\documentclass[aps,prl,twocolumn,showpacs,groupedaddress,nofootinbib]{revtex4}  
\usepackage{graphicx}  
\usepackage{dcolumn}   
\usepackage{bm}        
\usepackage{amssymb}   

\def\lsim{\mathrel{\rlap{\lower4pt\hbox{$\sim$}}
    \raise1pt\hbox{$<$}}}                

\newcommand{\ppb}{\mbox{\ensuremath{p\bar p}}}

\newcommand{\invpb}{pb$^{-1}$}

\newcommand{\met}{\mbox{\ensuremath{\,\slash\kern-.7emE_{T}}}}
\newcommand{\mht}{\mbox{\ensuremath{\,\slash\kern-.7emH_{T}}}}

\newcommand{\htj}{\mbox{\ensuremath{H_{T}}}}
\newcommand{\pt}{\mbox{\ensuremath{p_{T}}}}
\newcommand{\etadet}{\mbox{\ensuremath{\vert\eta_\mathrm{det}\vert}}}

\newcommand{\dphimin}{\mbox{\ensuremath{\Delta\Phi_\mathrm{min}}}}
\newcommand{\dphimax}{\mbox{\ensuremath{\Delta\Phi_\mathrm{max}}}}
\newcommand{\dphiadd}{\mbox{\ensuremath{\dphimax+\dphimin}}}

\newcommand{\st}{\mbox{\ensuremath{\tilde t}}}
\newcommand{\mst}{\mbox{\ensuremath{m_{\tilde t}}}}
\newcommand{\chiz}{\mbox{\ensuremath{\tilde\chi^0_1}}}
\newcommand{\mchi}{\mbox{\ensuremath{m_{\tilde\chi^0_1}}}}

\begin{document}


\hspace{5.2in} \mbox{Fermilab-Pub-06/396-E}

\title{Search for the pair production of scalar top quarks in the acoplanar 
charm jet \\ final state in \ppb\ collisions at $\sqrt{s}=1.96$\,TeV}

%
\author{                                                                      
V.M.~Abazov,$^{35}$                                                           
B.~Abbott,$^{75}$                                                             
M.~Abolins,$^{65}$                                                            
B.S.~Acharya,$^{28}$                                                          
M.~Adams,$^{51}$                                                              
T.~Adams,$^{49}$                                                              
E.~Aguilo,$^{5}$                                                              
S.H.~Ahn,$^{30}$                                                              
M.~Ahsan,$^{59}$                                                              
G.D.~Alexeev,$^{35}$
% ...
\\                                                                            
\vskip 0.30cm                                                                 
\centerline{(D\O\ Collaboration)}                                             
\vskip 0.30cm                                                                 
}                                                                             
\affiliation{                                                                 
\centerline{$^{1}$Universidad de Buenos Aires, Buenos Aires, Argentina}       
\centerline{$^{2}$LAFEX, Centro Brasileiro de Pesquisas F{\'\i}sicas,         
                  Rio de Janeiro, Brazil}                                     
\centerline{$^{3}$Universidade do Estado do Rio de Janeiro,                   
                  Rio de Janeiro, Brazil}                                     
\centerline{$^{4}$Instituto de F\'{\i}sica Te\'orica, Universidade            
                  Estadual Paulista, S\~ao Paulo, Brazil}                     
\centerline{$^{5}$University of Alberta, Edmonton, Alberta, Canada,           
                  Simon Fraser University, Burnaby, British Columbia, Canada,}
\centerline{York University, Toronto, Ontario, Canada, and                    
                  McGill University, Montreal, Quebec, Canada}                
\centerline{$^{6}$University of Science and Technology of China, Hefei,       
                  People's Republic of China}                                 
\centerline{$^{7}$Universidad de los Andes, Bogot\'{a}, Colombia}             
\centerline{$^{8}$Center for Particle Physics, Charles University,            
                  Prague, Czech Republic}                                     
\centerline{$^{9}$Czech Technical University, Prague, Czech Republic}         
\centerline{$^{10}$Center for Particle Physics, Institute of Physics,         
                   Academy of Sciences of the Czech Republic,                 
                   Prague, Czech Republic}                                    
\centerline{$^{11}$Universidad San Francisco de Quito, Quito, Ecuador}        
\centerline{$^{12}$Laboratoire de Physique Corpusculaire, IN2P3-CNRS,         
                   Universit\'e Blaise Pascal, Clermont-Ferrand, France}      
\centerline{$^{13}$Laboratoire de Physique Subatomique et de Cosmologie,      
                   IN2P3-CNRS, Universite de Grenoble 1, Grenoble, France}    
\centerline{$^{14}$CPPM, IN2P3-CNRS, Universit\'e de la M\'editerran\'ee,     
                   Marseille, France}                                         
\centerline{$^{15}$IN2P3-CNRS, Laboratoire de l'Acc\'el\'erateur              
                   Lin\'eaire, Orsay, France}                                 
\centerline{$^{16}$LPNHE, IN2P3-CNRS, Universit\'es Paris VI and VII,         
                   Paris, France}                                             
\centerline{$^{17}$DAPNIA/Service de Physique des Particules, CEA, Saclay,    
                   France}                                                    
\centerline{$^{18}$IPHC, IN2P3-CNRS, Universit\'e Louis Pasteur, Strasbourg,  
                    France, and Universit\'e de Haute Alsace,                 
                    Mulhouse, France}                                         
\centerline{$^{19}$Institut de Physique Nucl\'eaire de Lyon, IN2P3-CNRS,      
                   Universit\'e Claude Bernard, Villeurbanne, France}         
\centerline{$^{20}$III. Physikalisches Institut A, RWTH Aachen,               
                   Aachen, Germany}                                           
\centerline{$^{21}$Physikalisches Institut, Universit{\"a}t Bonn,             
                   Bonn, Germany}                                             
\centerline{$^{22}$Physikalisches Institut, Universit{\"a}t Freiburg,         
                   Freiburg, Germany}                                         
\centerline{$^{23}$Institut f{\"u}r Physik, Universit{\"a}t Mainz,            
                   Mainz, Germany}                                            
\centerline{$^{24}$Ludwig-Maximilians-Universit{\"a}t M{\"u}nchen,            
                   M{\"u}nchen, Germany}                                      
\centerline{$^{25}$Fachbereich Physik, University of Wuppertal,               
                   Wuppertal, Germany}                                        
\centerline{$^{26}$Panjab University, Chandigarh, India}                      
\centerline{$^{27}$Delhi University, Delhi, India}                            
\centerline{$^{28}$Tata Institute of Fundamental Research, Mumbai, India}     
\centerline{$^{29}$University College Dublin, Dublin, Ireland}                
\centerline{$^{30}$Korea Detector Laboratory, Korea University,               
                   Seoul, Korea}                                              
\centerline{$^{31}$SungKyunKwan University, Suwon, Korea}                     
\centerline{$^{32}$CINVESTAV, Mexico City, Mexico}                            
\centerline{$^{33}$FOM-Institute NIKHEF and University of                     
                   Amsterdam/NIKHEF, Amsterdam, The Netherlands}              
\centerline{$^{34}$Radboud University Nijmegen/NIKHEF, Nijmegen, The          
                  Netherlands}                                                
\centerline{$^{35}$Joint Institute for Nuclear Research, Dubna, Russia}       
\centerline{$^{36}$Institute for Theoretical and Experimental Physics,        
                   Moscow, Russia}                                            
\centerline{$^{37}$Moscow State University, Moscow, Russia}                   
\centerline{$^{38}$Institute for High Energy Physics, Protvino, Russia}       
\centerline{$^{39}$Petersburg Nuclear Physics Institute,                      
                   St. Petersburg, Russia}                                    
\centerline{$^{40}$Lund University, Lund, Sweden, Royal Institute of          
                   Technology and Stockholm University, Stockholm,            
                   Sweden, and}                                               
\centerline{Uppsala University, Uppsala, Sweden}                              
\centerline{$^{41}$Physik Institut der Universit{\"a}t Z{\"u}rich,            
                   Z{\"u}rich, Switzerland}                                   
\centerline{$^{42}$Lancaster University, Lancaster, United Kingdom}           
\centerline{$^{43}$Imperial College, London, United Kingdom}                  
\centerline{$^{44}$University of Manchester, Manchester, United Kingdom}      
\centerline{$^{45}$University of Arizona, Tucson, Arizona 85721, USA}         
\centerline{$^{46}$Lawrence Berkeley National Laboratory and University of    
                   California, Berkeley, California 94720, USA}               
\centerline{$^{47}$California State University, Fresno, California 93740, USA}
\centerline{$^{48}$University of California, Riverside, California 92521, USA}
\centerline{$^{49}$Florida State University, Tallahassee, Florida 32306, USA} 
\centerline{$^{50}$Fermi National Accelerator Laboratory,                     
            Batavia, Illinois 60510, USA}                                     
\centerline{$^{51}$University of Illinois at Chicago,                         
            Chicago, Illinois 60607, USA}                                     
\centerline{$^{52}$Northern Illinois University, DeKalb, Illinois 60115, USA} 
\centerline{$^{53}$Northwestern University, Evanston, Illinois 60208, USA}    
\centerline{$^{54}$Indiana University, Bloomington, Indiana 47405, USA}       
\centerline{$^{55}$University of Notre Dame, Notre Dame, Indiana 46556, USA}  
\centerline{$^{56}$Purdue University Calumet, Hammond, Indiana 46323, USA}    
\centerline{$^{57}$Iowa State University, Ames, Iowa 50011, USA}              
\centerline{$^{58}$University of Kansas, Lawrence, Kansas 66045, USA}         
\centerline{$^{59}$Kansas State University, Manhattan, Kansas 66506, USA}     
\centerline{$^{60}$Louisiana Tech University, Ruston, Louisiana 71272, USA}   
\centerline{$^{61}$University of Maryland, College Park, Maryland 20742, USA} 
\centerline{$^{62}$Boston University, Boston, Massachusetts 02215, USA}       
\centerline{$^{63}$Northeastern University, Boston, Massachusetts 02115, USA} 
\centerline{$^{64}$University of Michigan, Ann Arbor, Michigan 48109, USA}    
\centerline{$^{65}$Michigan State University,                                 
            East Lansing, Michigan 48824, USA}                                
\centerline{$^{66}$University of Mississippi,                                 
            University, Mississippi 38677, USA}                               
\centerline{$^{67}$University of Nebraska, Lincoln, Nebraska 68588, USA}      
\centerline{$^{68}$Princeton University, Princeton, New Jersey 08544, USA}    
\centerline{$^{69}$State University of New York, Buffalo, New York 14260, USA}
\centerline{$^{70}$Columbia University, New York, New York 10027, USA}        
\centerline{$^{71}$University of Rochester, Rochester, New York 14627, USA}   
\centerline{$^{72}$State University of New York,                              
            Stony Brook, New York 11794, USA}                                 
\centerline{$^{73}$Brookhaven National Laboratory, Upton, New York 11973, USA}
\centerline{$^{74}$Langston University, Langston, Oklahoma 73050, USA}        
\centerline{$^{75}$University of Oklahoma, Norman, Oklahoma 73019, USA}       
\centerline{$^{76}$Oklahoma State University, Stillwater, Oklahoma 74078, USA}
\centerline{$^{77}$Brown University, Providence, Rhode Island 02912, USA}     
\centerline{$^{78}$University of Texas, Arlington, Texas 76019, USA}          
\centerline{$^{79}$Southern Methodist University, Dallas, Texas 75275, USA}   
\centerline{$^{80}$Rice University, Houston, Texas 77005, USA}                
\centerline{$^{81}$University of Virginia, Charlottesville,                   
            Virginia 22901, USA}                                              
\centerline{$^{82}$University of Washington, Seattle, Washington 98195, USA}  
}                                                                             
\date{November 1, 2006}

\begin{abstract}
A search for the pair production of scalar top quarks, \st, has been performed 
in 360\,\invpb\ of data from \ppb\ collisions at a center-of-mass energy of 
1.96\,TeV, collected by the D0 detector at the Fermilab Tevatron collider. 
The \st\ decay mode considered is $\st\to c\chiz$, where $\chiz$ is the 
lightest 
supersymmetric particle. The topology analyzed therefore consists of a pair of 
acoplanar heavy-flavor jets with missing transverse energy. 
The data and standard model expectation are in agreement, 
and a 95\% C.L. exclusion domain in the 
(\mst,\mchi) plane has been determined, extending the domain excluded by
previous experiments.
\end{abstract}

\maketitle 

Supersymmetric (SUSY) models\,\cite{HK} predict the existence of new particles,
carrying the same quantum numbers as their standard model (SM) partners, but 
differing by half a unit of spin. For instance, there are two scalar-quark 
fields associated with the left- and right-handed degrees of freedom of each 
ordinary quark. The mass eigenstates result from the diagonalization of a mass 
matrix, with elements determined by the specific SUSY-breaking pattern. 
A light SUSY partner of the top quark, or stop, is a generic prediction of 
models in which the scalar quark masses are equal at the grand unification 
scale. A first reason is 
that, due to the impact of the large top quark Yukawa coupling in the 
renormalization group equations, the diagonal elements of the mass matrix are 
driven to values smaller than those for the other scalar quarks at the 
electroweak scale \,\cite{BBO}. A second reason is that the off-diagonal terms 
are proportional to the relevant quark mass, and hence are much larger in the 
case of the top quark. The mass eigenstates are therefore broadly split, with 
the mass of the lighter stop \st\ thus driven to an even lower 
value\,\cite{EsRz}. Finally, a light stop is a necessary ingredient in the 
context of electroweak baryogenesis\,\cite{CQW}.

In models with $R$-parity conservation\,\cite{Fayet}, the lightest SUSY 
particle (LSP) is stable, and cosmological constraints imply that it should be 
neutral and colorless\,\cite{cosmo}. In a large class of SUSY models, the 
lightest of the neutralinos --- the mass eigenstates resulting from 
the mixing of the SUSY partners of the neutral gauge and Higgs bosons ---
is the LSP, which furthermore appears as a viable dark matter
candidate. 
In the following, it will be assumed that $R$-parity is conserved and 
that the LSP is the lightest neutralino $\chiz$.

The dominant stop decay modes are expected to be $\st\to t\chiz$ and 
$\st\to b\tilde\chi^+_1$, where the chargino $\tilde\chi^+_1$ is the lighter 
of the two mass 
eigenstates resulting from the mixing of the SUSY partners of the charged 
gauge and Higgs bosons. However, in the \st\ mass range of interest in this 
Letter, the $\st\to t\chiz$ decay mode is kinematically forbidden. In the
following, the region of SUSY parameter space with 
$\mst < m_b+m_{\tilde\chi^+_1}$ and $\mst <  M_W+m_b+\mchi$
is considered, and it is assumed that 
$\st\to c\chiz$ , a flavor-changing loop 
decay\,\cite{HiKo}, is the only relevant decay mode, i.e., that the  
tree-level four-body decays\,\cite{BDM} $\st\to bf\bar f'\chiz$ 
can be neglected.

In $p\bar p$ collisions, stop pair production proceeds via $q\bar q$ 
annihilation and gluon-gluon fusion. The cross section has very little 
dependence on SUSY parameters other than the stop mass. At the center-of-mass 
energy of 1.96\,TeV available in Run~II of the Fermilab Tevatron collider, it 
ranges from 15 to 2.25\,pb for stop masses from 100 to 140\,GeV, as calculated 
at next-to-leading order (NLO) with {\sc prospino}\,\cite{PROSPINO}, for equal
renormalization and factorization scales $\mu_{rf}=\mst$ and using the
CTEQ6.1M parton distribution functions (PDFs)\,\cite{CTEQ6}. The final state
topology resulting from the $\st\to c\chiz$ decay is a pair of acoplanar jets,
with large missing transverse energy \met\ carried away by the two weakly
interacting LSPs. Previous searches in this topology performed at LEP
excluded stop masses smaller than $\approx 100$\,GeV, 
essentially independent of
the stop-$\chiz$ mass difference\,\cite{LEP}. Searches in data from
the Run~I of the
Tevatron\,\cite{CDF,DRunI} extended the domain excluded at LEP to larger stop 
masses, but for \chiz\ masses not exceeding $\approx 50$\,GeV. The largest stop
mass excluded was 122\,GeV, for $\mchi=45$\,GeV\,\cite{DRunI}. In this Letter,
we report on a similar search, performed in data collected using the D0 
detector during Run~II of the Tevatron.

The acoplanar jet topology may arise from new physics processes other than stop
pair production. Recently, the D0 Collaboration performed a search for pair
production of leptoquarks decaying into a quark and a neutrino\,\cite{lepto}, 
which leads to the same topology. The analysis reported here is largely based
on that leptoquark search. In the following, only a brief summary of the 
common aspects is given, while the specific features relevant for 
the stop search are presented in greater detail. The main differences arise
from the LSP mass, which leads to smaller jet transverse energies and to a 
reduced \met, compared to the case of leptoquark decays which involve 
nearly massless 
neutrinos. Another characteristic feature of stop decays is that charm jets 
are produced, while first-generation leptoquarks decay to light-flavor jets.

A thorough description of the D0 detector can be found in Ref.~\cite{detector}.
The central tracking system consists of a silicon microstrip tracker and a 
fiber tracker, both located within a 2\,T superconducting solenoidal magnet. 
A liquid-argon and uranium calorimeter covers pseudorapidities 
$|\eta| \lsim 4.2$, 
where $\eta=-\ln \left[ \tan \left( \theta/2 \right) \right]$ and $\theta$ 
is the polar angle with respect to the proton beam direction. 
An outer muon system, covering $|\eta|<2$, consists of layers of tracking 
detectors and scintillation counters on both sides of 1.8~T iron toroids. 

For this search, $\approx 14$ million events collected from April 2003 to 
August 2004 with a \mbox{jets + $\met$} trigger were analyzed, corresponding 
to an integrated luminosity\,\footnote{This value differs from
the one used in Ref.\,\cite{lepto} due to a recent adjustment of the D0
luminosity constant\,\cite{lumi}.} of 360\,\invpb.
The offline analysis utilized jets reconstructed with the iterative midpoint 
cone algorithm\,\cite{jetalgo} with a cone size of 0.5.
Only jets with transverse momentum
$\pt > 15$\,GeV were considered in the analysis. 
The \met\ was calculated using all calorimeter cells, corrected for
the energy calibration of reconstructed jets, as determined from the 
transverse momentum balance in photon+jet events, and for the momentum of
reconstructed muons. 

Signal efficiencies and SM backgrounds were evaluated 
using a full {\sc geant-3}\,\cite{GEANT} based simulation of events, with a 
Poisson average of 0.8 minimum-bias events superimposed, corresponding to the
luminosity profile of the data sample analyzed. These simulated events were
reconstructed in the same way as the data.
In the bulk of events from QCD multijet production, no significant \met\ is 
expected. Jet energy mismeasurements due to the limited detector resolution 
may however lead to large measured \met\ values. 
This ``instrumental background'' was not simulated, and its contribution
estimated directly from the data. 
In the following, ``standard model (SM) 
background'' stands for ``non-QCD standard model (SM) background.''
Leptonic $W$ decays, as well
as $Z\to\nu\nu$ are sources of energetic neutrinos, hence of genuine \met.
The SM processes expected to yield the largest background contributions are 
therefore vector boson production in association with jets. 
They were generated with 
{\sc alpgen 1.3}\,\cite{ALPGEN}, interfaced 
with {\sc pythia 6.202}\,\cite{PYTHIA} for the simulation of initial and final 
state radiation and for jet hadronization. 
The PDFs used were CTEQ5L\,\cite{PDF5}.
The NLO cross sections for vector boson production in
association with jets were calculated with {\sc mcfm 3.4.4}\,\cite{MCFM}.
Vector-boson pair, $t\bar t$, and single top quark production were also 
considered. 
Signal samples of 10\,000 events were generated with {\sc pythia} and the 
CTEQ5L PDFs for stop 
masses ranging from 95 to 145\,GeV and for $\chiz$ masses from 40 to 70\,GeV, 
both in steps of 5\,GeV.

The following selection criteria were applied, independent of the stop and 
$\chiz$ masses:
there had to be at least two jets; 
the vector sum \mht\ of all jet transverse momenta 
($\mht = \vert\sum_{\text{jets}} \overrightarrow p_T\vert$)
as well as the missing transverse energy had to exceed 40\,GeV;
the leading and subleading jets (where jets are ordered according to their 
transverse momentum) had to be central ($\etadet < 1.5$, 
where $\eta_\mathrm{det}$ is the pseudorapidity measured from the detector 
center), with transverse momenta exceeding 40 and 20\,GeV, respectively, and
they had to be confirmed by charged particle tracks\,\cite{lepto};
the acoplanarity $\Delta\Phi$ of the two leading jets had to be smaller
than $165^\circ$, where $\Delta\Phi$ is the difference between the two jet 
azimuthal angles; 
the longitudinal position of the primary vertex had to be less than 
60\,cm away from the center of the detector.
At this point, 99\,884 events were selected, largely dominated by instrumental
background from multijet events. The efficiency for a reference signal with
$\mst=140$\,GeV and $\mchi=60$\,GeV was 30\%. 

The jet multiplicity 
distribution revealed that most of the selected events contained at least 
three jets, due to the acoplanarity requirement. Therefore, only events 
containing exactly two jets were retained, leaving 27\,853 data events
with an efficiency of 22\% for the reference signal. The inefficiency 
associated with the rejection of events with more than two 
jets was evaluated, based on studies of jet multiplicities in real and 
simulated $Z\to ee$ events with at least two jets, 
where the two leading jets fulfilled similar selection
criteria as in the analysis. This study also showed that the kinematic 
variables used in the analysis were adequately simulated.
Standard model backgrounds
from $W\to\ell\nu$+jet processes were greatly reduced by requiring that there 
be no isolated electron or muon with $\pt>10$\,GeV, and no isolated charged 
particle track with $\pt>5$\,GeV\,\cite{lepto}. This retained 22\,106 data 
events, with an efficiency of 19\% for the reference signal. 

\begin{figure*}[htbp]
\begin{tabular}{cc}
\includegraphics[width=8.5cm]{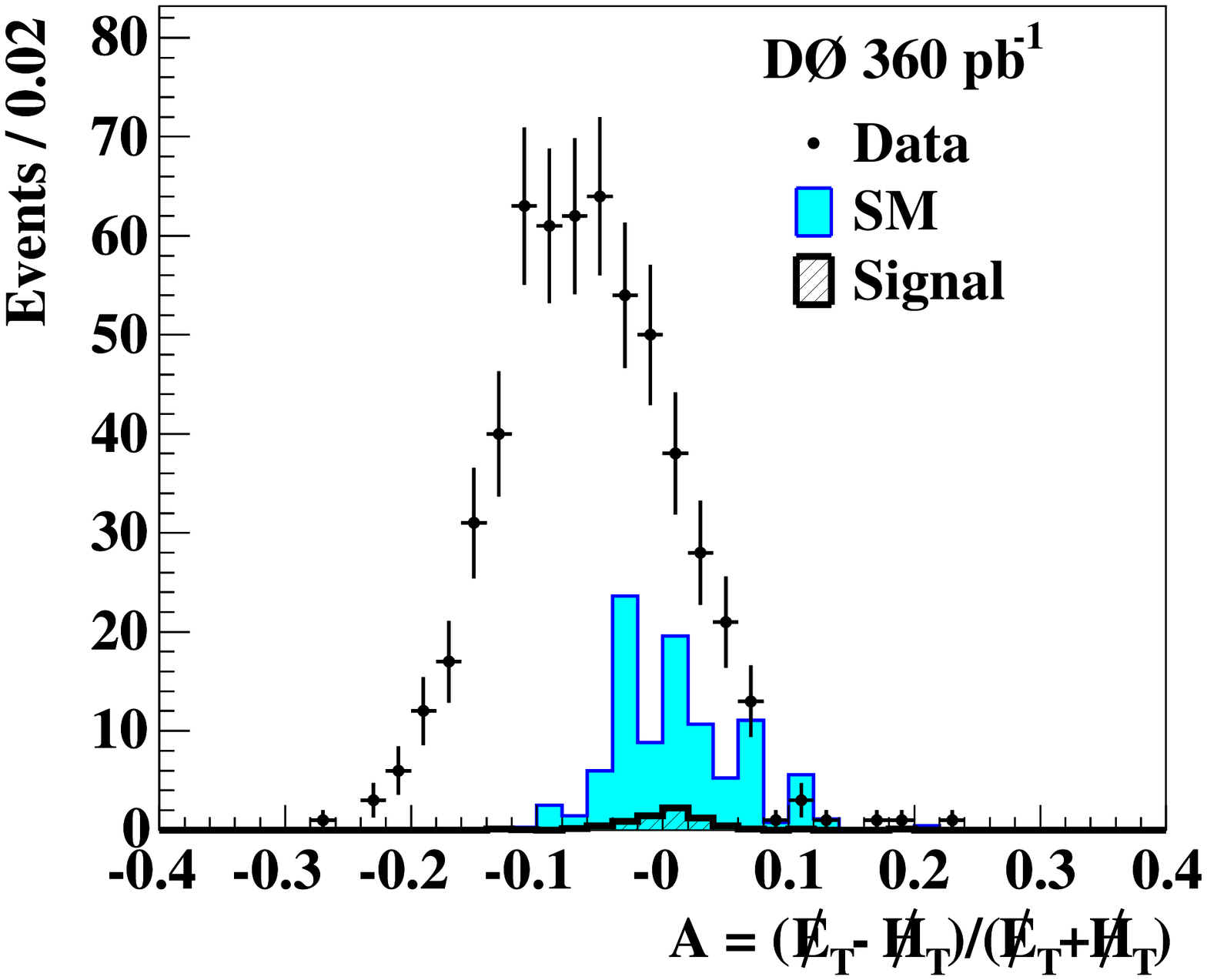} &
\includegraphics[width=8.5cm]{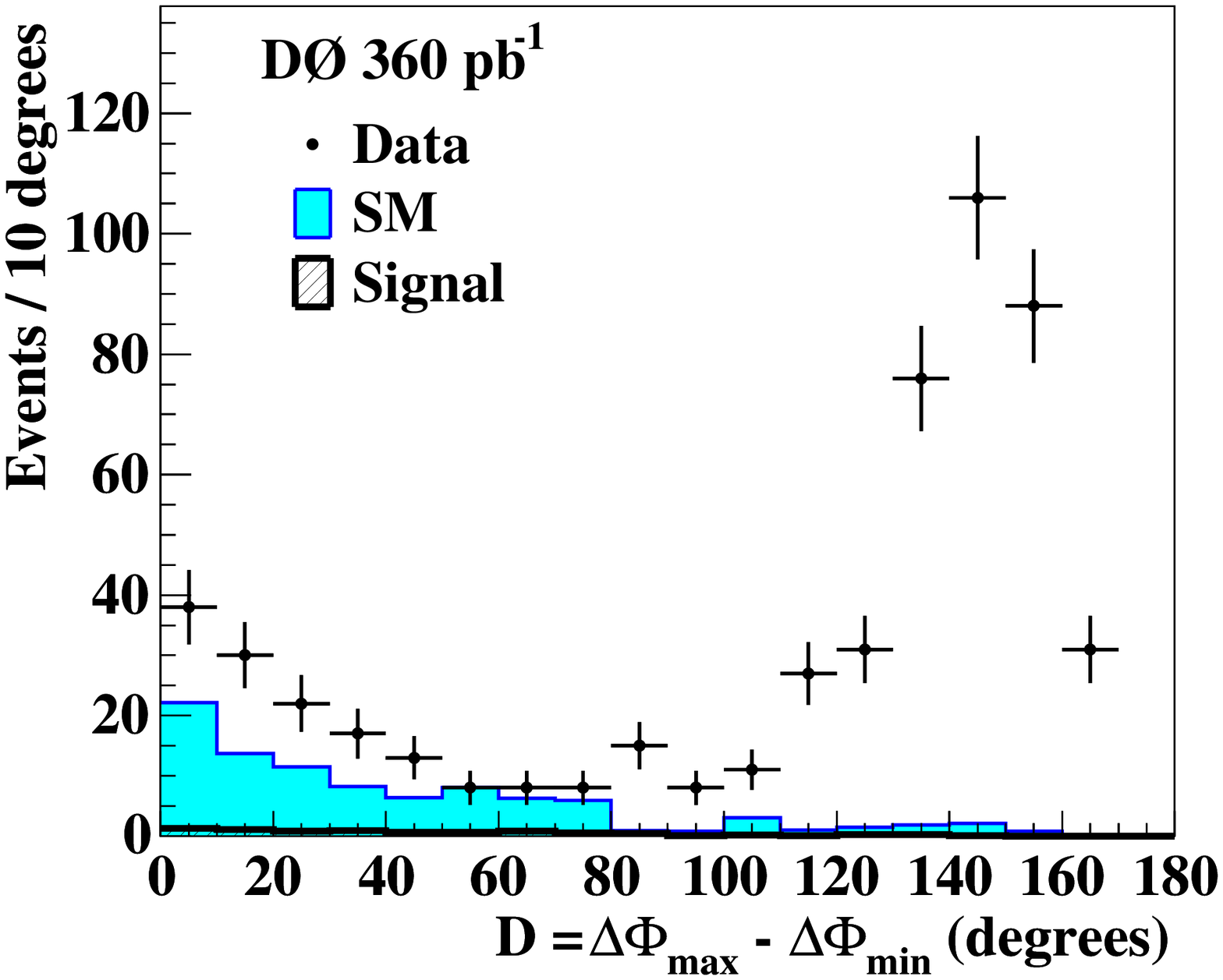} \\
\includegraphics[width=8.5cm]{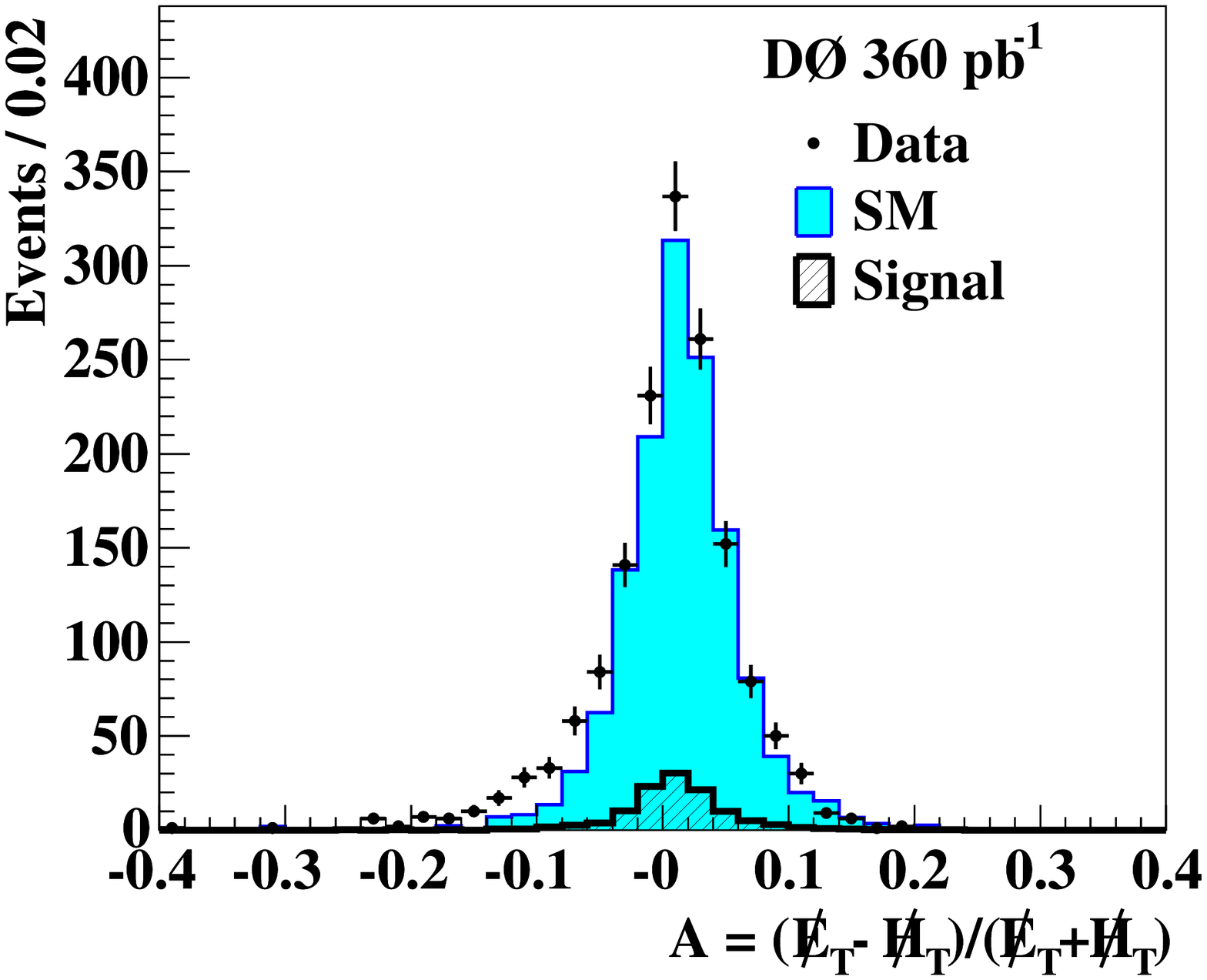} &
\includegraphics[width=8.5cm]{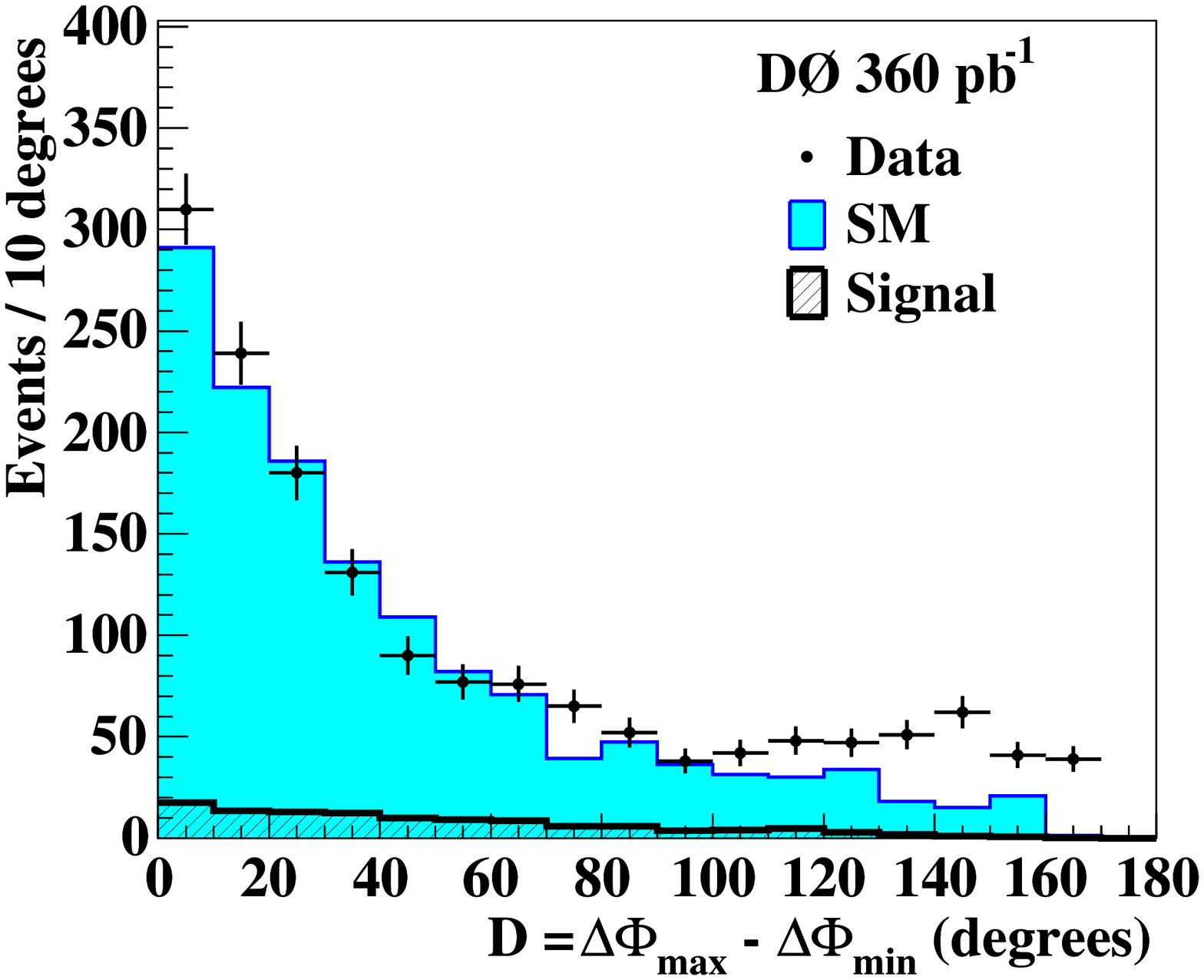}
\end{tabular}{cc}
\caption{\label{makdphi}
Distributions of the asymmetry ${\cal A}=(\met-\mht)/(\met+\mht)$ 
with the cut on 
${\cal D}=\dphimax-\dphimin$ inverted (top-left) or applied (bottom-left) 
and of ${\cal D}$ 
with the cut on 
${\cal A}$ inverted (top-right) or applied (bottom-right)
for data (points with error bars), 
for SM backgrounds (filled histogram), and for a signal with
$\mst=140$\,GeV and $\mchi=60$\,GeV (hatched histogram).  
The \met\ cut at 60\,GeV has been applied. In the bottom plots, 
the excesses in data 
for ${\cal A} < -0.05$ and for ${\cal D} > 120^\circ$
are attributed to the residual non-simulated instrumental background. 
}
\end{figure*}

Most of the remaining instrumental background was eliminated by 
the following requirements. The \met\ had to exceed 60\,GeV, and the
difference ${\cal D}=\dphimax-\dphimin$ had to be smaller than $120^\circ$, 
where \dphimin\ and \dphimax\ are the minimum and maximum of the azimuthal 
angles between the \met\ direction and the directions of the two jets, 
respectively. These criteria take advantage of the facts that, for the 
instrumental background, the \met\ distribution is steeply decreasing, and its
direction tends to be close to that of a mismeasured jet. In addition,
the asymmetry ${\cal A}=(\met-\mht)/(\met+\mht)$ was required to be larger 
than $-0.05$. This 
variable is sensitive to the amount of energy deposited in the calorimeter
that was not clustered into jets. It can be seen in Fig.\,\ref{makdphi} that
both ${\cal D}$ and ${\cal A}$ are effective in discriminating SM 
backgrounds and signal from the instrumental background. After these
requirements, 1\,348 data events were retained, while 1\,292 $\pm$ 45 events
were expected from SM backgrounds, where the uncertainty is statistical. The
efficiency for the reference signal was 13\%. 
There was no evidence at this point for any significant 
instrumental background remaining. This background has therefore been 
neglected in the following.

To increase the search sensitivity, advantage was then taken of the presence 
of charm jets in the signal. A lifetime-based heavy-flavor tagging algorithm 
was used for this purpose, which involves a probability built from the 
impact parameter significances of the tracks belonging to a jet\,\cite{JLIP}. 
The impact parameter of a track is its distance of closest approach to the 
event vertex, in a plane perpendicular to the beam axis, and the significance 
is obtained by normalization to the impact parameter uncertainty. This 
probability is constructed such that its distribution is uniform for 
light-flavor jets and peaks towards zero for heavy-flavor jets. 
In order to cope with differences in track reconstruction efficiencies in data 
and in simulation, the heavy-flavor tagging algorithm was applied directly 
only to the data, while flavor-dependent tagging probabilities measured in 
dedicated data samples
were applied to the simulated jets. The probability cut used in this 
analysis was such that typically 4\% of the light-flavor jets were tagged
(central jets with $\pt\approx 50$\,GeV).
The corresponding typical tagging efficiencies for $c$ and $b$ quark jets were 
30\% and 65\%, respectively. Jets resulting from $\tau$ decays were tagged with
a typical efficiency of 20\%. By requiring that at least one jet be tagged, 
183 data events were selected, while $186 \pm 16$ SM background events 
were expected, where the uncertainty is statistical. 
The efficiency for the reference signal was 6.5\%. 

\begin{figure}[htbp]
\includegraphics[width=8.5cm]{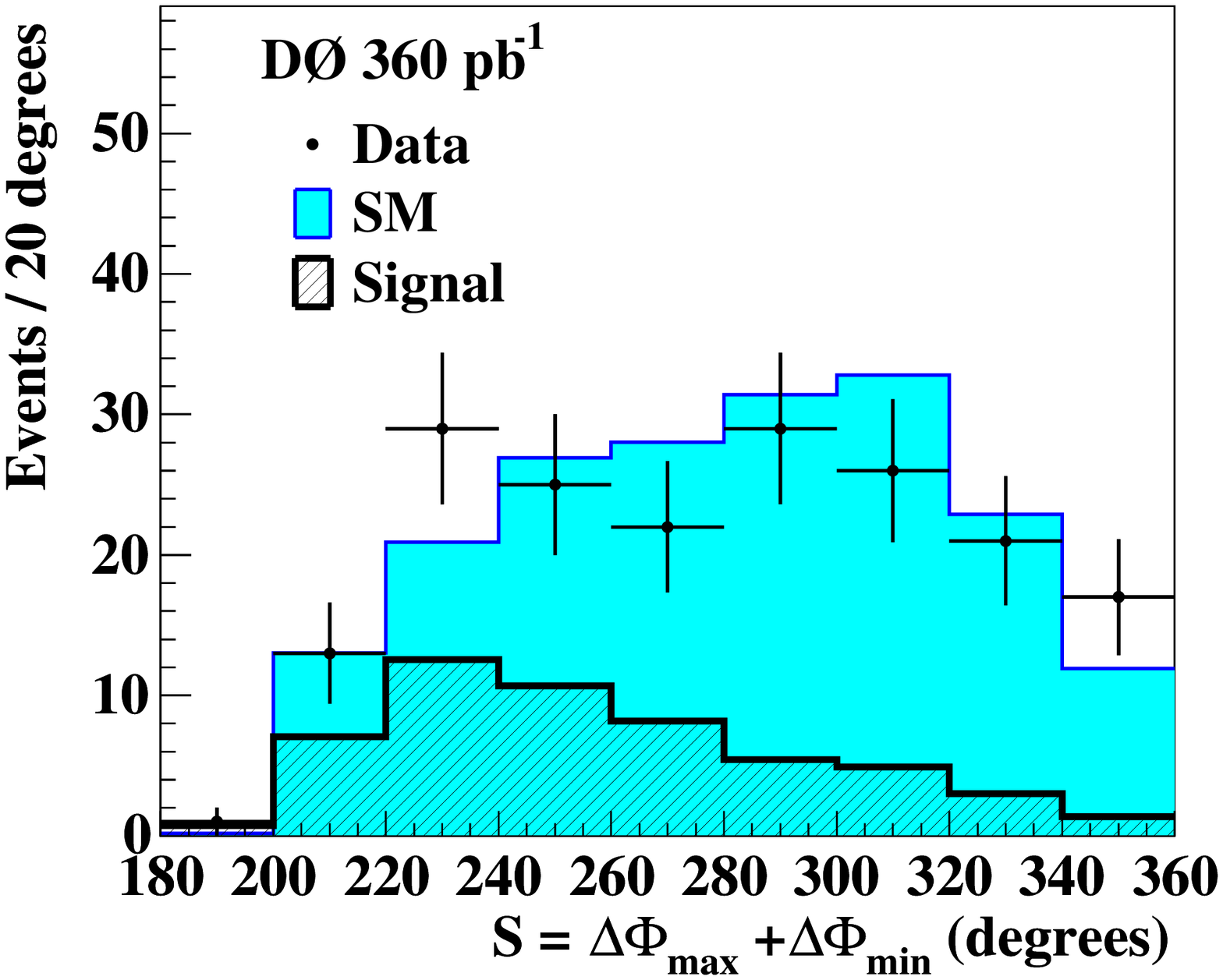} \\
\includegraphics[width=8.5cm]{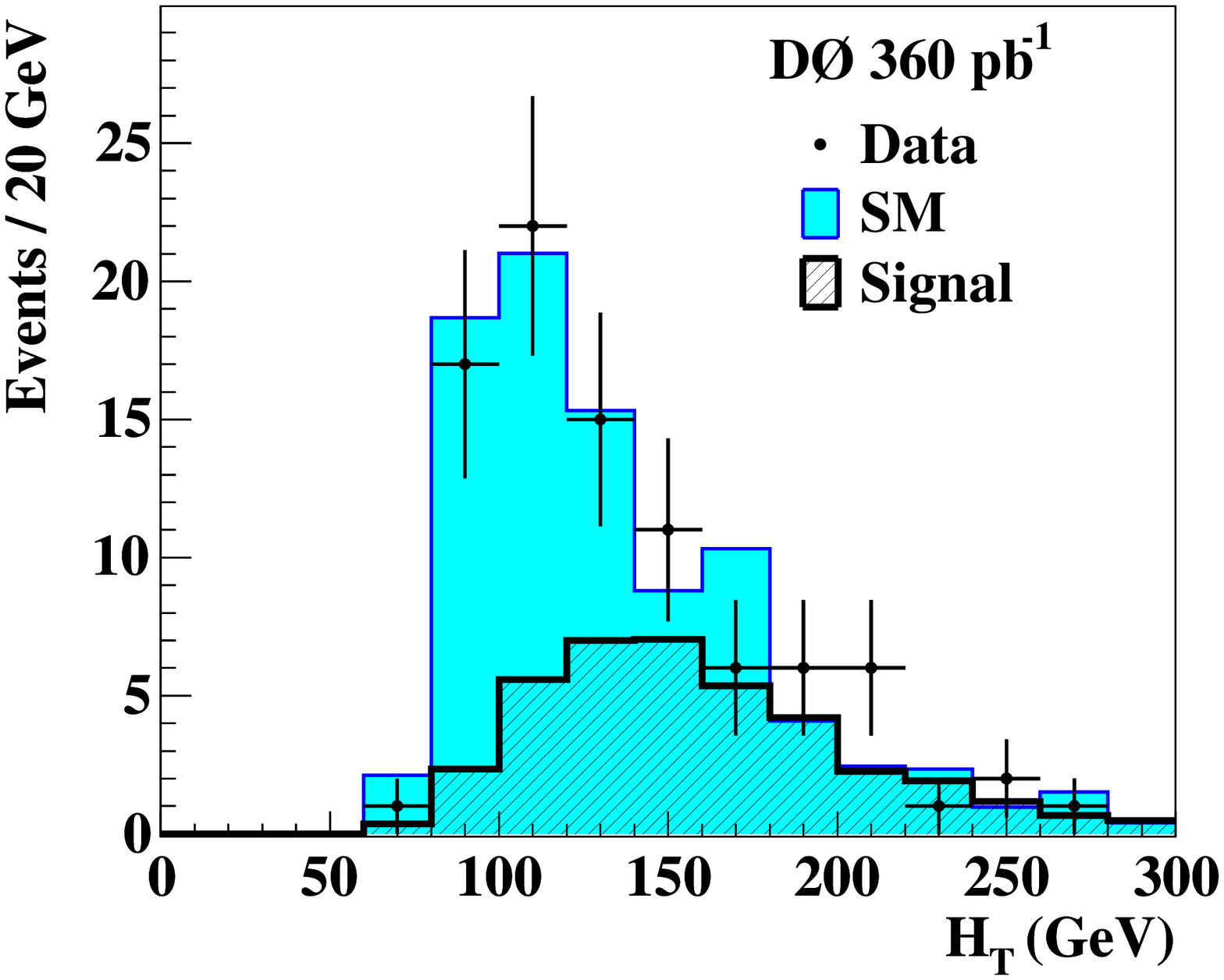}
\caption{\label{varoptim}
Distributions of ${\cal S}=\dphiadd$ before optimization (top), and of
\htj\ after optimization for $\mst=140$\,GeV but with the cut on \htj\ removed 
(bottom), for data (points with error bars),
for SM backgrounds (filled histogram), and for a signal with
$\mst=140$\,GeV and $\mchi=60$\,GeV (hatched histogram).}
\end{figure}

Since the signal topology depends on the stop and $\chiz$ masses, additional 
selection criteria on three kinematic variables were simultaneously optimized 
for each mass combination. These variables were the scalar sum 
$\htj = \sum_{\text{jets}} \vert\overrightarrow p_T\vert$ 
of the 
jet transverse momenta in steps of 20\,GeV, \met\ in steps of 
10\,GeV, and ${\cal S}=\dphiadd$ in steps of $10^\circ$. It can be seen in
Fig.\,\ref{varoptim} that this last variable provides good discrimination 
between signal and SM backgrounds. 
For \htj\ and 
\met, the selection retained events above the cut value, while for ${\cal S}$,
events below the cut value were selected. 
For each stop and \chiz\ mass combination tested, 
all sets of cuts were considered. 
For each set, 
the value $\langle \mbox{\sl CL}_s \rangle$ of the signal confidence level\,\cite{CLs} expected
if only background were present 
was computed, with the systematic uncertainties discussed below 
taken into account. For a given stop mass, 
the expected lower limit on \mchi\ was determined as the 
\chiz\ mass for which $\langle \mbox{\sl CL}_s \rangle = 5\%$, by interpolation across the \mchi\ values
tested. The set leading to the largest expected lower limit on \mchi\ was 
selected as the optimal one for the stop mass considered. 
In all cases, a \met\ cut at 60\,GeV was selected. The results of the 
optimization for the other variables
are given in Table\,\ref{resoptim}, together with the numbers of events 
selected in the data and expected from SM backgrounds. Signal efficiencies and
numbers of signal events expected are given in Table\,\ref{sigeff} for three 
mass combinations close to the edge of the sensitivity domain of the analysis. 

\begin{figure}[htbp]
\includegraphics[width=8.5cm]{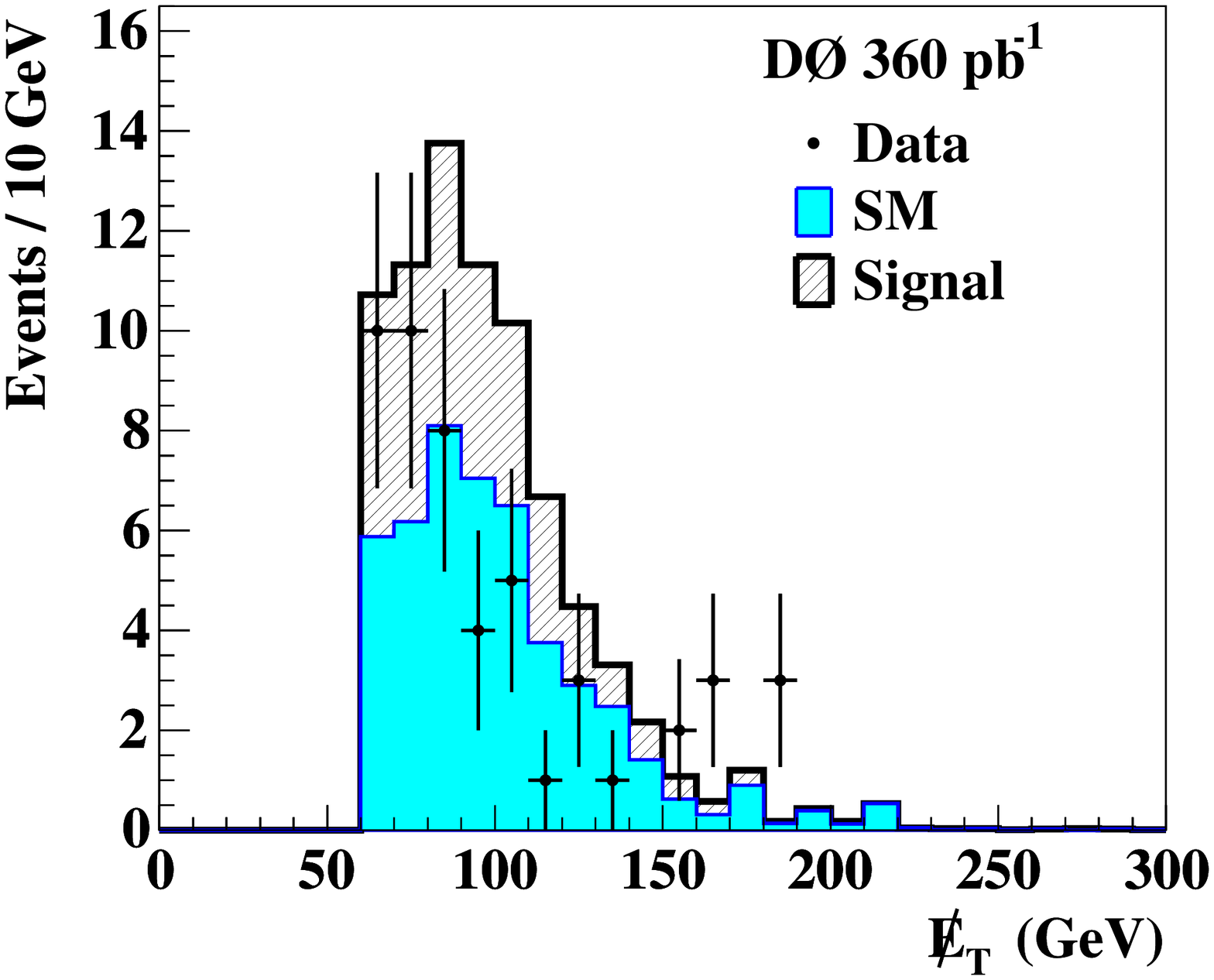}
\caption{\label{metend}
Final \met\ distribution for data (points with error bars), 
for SM backgrounds (filled histogram), and,  on top of the SM backgrounds,
for a signal with $\mst=140$\,GeV and $\mchi=60$\,GeV (hatched histogram).}
\end{figure}

The distribution of 
\htj\ shown in Fig.\,\ref{varoptim}  
and the final distribution of \met\ shown in 
Fig.\,\ref{metend}
were obtained after optimization for a 
stop mass of 140\,GeV.
An excess at large \met\ is observed in the data with respect to the
expectation: there are eight data events with \mbox{$\met>150$\,GeV}, 
while $3.2 \pm 1.4$ events are expected from SM backgrounds. A detailed 
scrutiny of those events was performed, that did not reveal any anomaly 
such as clusterings in some of the kinematic variables, signs of leptons
unidentified by the standard algorithms, heavy flavor tagging probabilities
different from what is observed in the rest of the selected events. The 
data taking conditions were also carefully checked for signs of detector 
malfunctions and visual scans were performed. It can also be noted that
such large \met\ values are beyond what is expected from a stop signal.

\begin{table}[htbp]
\renewcommand{\arraystretch}{1.4}
\caption{\label{resoptim}
Results of the optimization: stop mass range in GeV, \htj\ cut value in GeV, 
and ${\cal S}$ cut value in degrees. In all cases, a \met\ cut at 60\,GeV was
selected. The numbers of events observed and expected from 
SM backgrounds are also given; the first uncertainties are statistical, and the
second systematic.}
\begin{ruledtabular}
\begin{tabular}{|c|c|c|c|l|}
\mst & \htj & ${\cal S}$ & \# observed & \# expected \\
\hline
95 -- 115    & $>80$  & $<260$ & 68 & $59.9 \pm 9.6$ \mbox{$^{+11.7}_{-9.7}$} \\
120          & $>80$  & $<280$ & 89 & $86.4 \pm 11.3$ \mbox{$^{+16.2}_{-14.2}$} \\
125 -- 140   & $>120$ & $<280$ & 50 & $47.0 \pm 8.0$ \mbox{$^{+9.7}_{-7.9}$} \\
145          & $>120$ & $<300$ & 57 & $53.8 \pm 8.3$ \mbox{$^{+10.8}_{-9.2}$} \\
\end{tabular}
\end{ruledtabular}
\end{table}
  
\begin{table}[htbp]
\renewcommand{\arraystretch}{1.4}
\caption{\label{sigeff}
For three stop and \chiz\ mass combinations, in GeV, signal efficiencies 
(Eff.) and numbers of signal events expected, where the 
first uncertainties are statistical and the second systematic. The stop pair 
production cross section upper limits at 95\% C.L. are also given 
($\sigma_\mathrm{UL}$), as well as the NLO theoretical cross section 
($\sigma_\mathrm{Th}$), both in pb.}
\begin{ruledtabular}
\begin{tabular}{|c|c|c|c|c|}
(\mst,\mchi) & Eff. (\%) & \# expected & $\sigma_\mathrm{UL}$ & 
$\sigma_\mathrm{Th}$ \\
\hline
(100,55) & 0.75 & $40.4 \pm 4.6$ \mbox{$^{+5.3}_{-5.4}$} & 15.8 & 15.0 \\
(120,65) & 2.04 & $40.0 \pm 2.8$ \mbox{$^{+5.6}_{-5.2}$} & 6.57 & 5.43 \\
(140,60) & 3.74 & $30.3 \pm 1.6$ \mbox{$^{+4.8}_{-5.3}$} & 2.38 & 2.25 \\
\end{tabular}
\end{ruledtabular}
\end{table}

The SM background composition is detailed in Table\,\ref{smbg} for the 
selection optimized for $\mst = 140$\, GeV. 
As expected, the largest contributions come
from ($Z\to\nu\nu$ and $W\to\ell\nu$)+light-flavor jets. This is due to the 
loose heavy-flavor tagging criterion which was selected in order to be 
efficient for charm jets. Vector boson production with heavy-flavor jets
gives rather small contributions because of the comparatively small cross 
sections. 

\begin{table}
\caption{\label{smbg} Numbers of events expected from the various SM background
processes in the
selection optimized for $\mst = 140$\, GeV. The uncertainties are statistical. 
In the vector boson + jets backgrounds, ``jet'' stands for ``light-flavor 
jet.''}
\begin{ruledtabular}
\begin{tabular}{|lr|}
SM process & \# expected \\
\hline
$Z\to\nu\nu$+jets                    & $13.9 \pm 2.8$ \\
$Z\to\nu\nu$+$c\bar c$               & $1.7 \pm 0.4$ \\
$Z\to\nu\nu$+$b\bar b$               & $3.5 \pm 0.2$ \\
$W\to\ell\nu$+jets                   & $19.5 \pm 7.4$ \\
$W\to\ell\nu$+($c\bar c$ or $c$+jet) & $1.8 \pm 0.5$ \\
$W\to\ell\nu$+$b\bar b$              & $1.5 \pm 0.2$ \\
$t\bar t$ and single top             & $4.1 \pm 0.2$ \\
$WW$, $WZ$, $ZZ$                     & $1.1 \pm 0.2 $ \\
\hline
Total & $47.0 \pm 8.0$ \\
\end{tabular}
\end{ruledtabular}
\end{table}

Systematic uncertainties were evaluated for each combination of stop and
$\chiz$ masses, according to the corresponding optimized selection criteria.
They are listed below for the reference signal. The following are fully 
correlated between SM-background and signal expectations:
from the jet energy calibration and resolution, 
                           $^{+13}_{-6}$\% for the SM background and 
                           $^{+3}_{-4}$\% for the signal;
from the jet multiplicity cut, 3\%;
from the trigger efficiency, 2\% after all selection cuts;
from the heavy-flavor tagging, 6\% for the SM background and 
                               7\% for the signal;
from the integrated luminosity of the analysis sample, 6\%.
In addition to the 17\% statistical uncertainty of the simulation, 
the normalization of the SM background expectation carries a 13\% 
uncertainty, as inferred from a comparison of data and simulated 
$(Z\to ee)+2$-jet events. 
The statistical uncertainty of the signal simulation is 5\%.
Finally, the uncertainty on the signal efficiency due to the PDF choice was 
determined to be $^{+6}_{-4}$\%, using the CTEQ6.1M error set\,\cite{CTEQ6}.

\begin{figure}
\includegraphics[width=8.5cm]{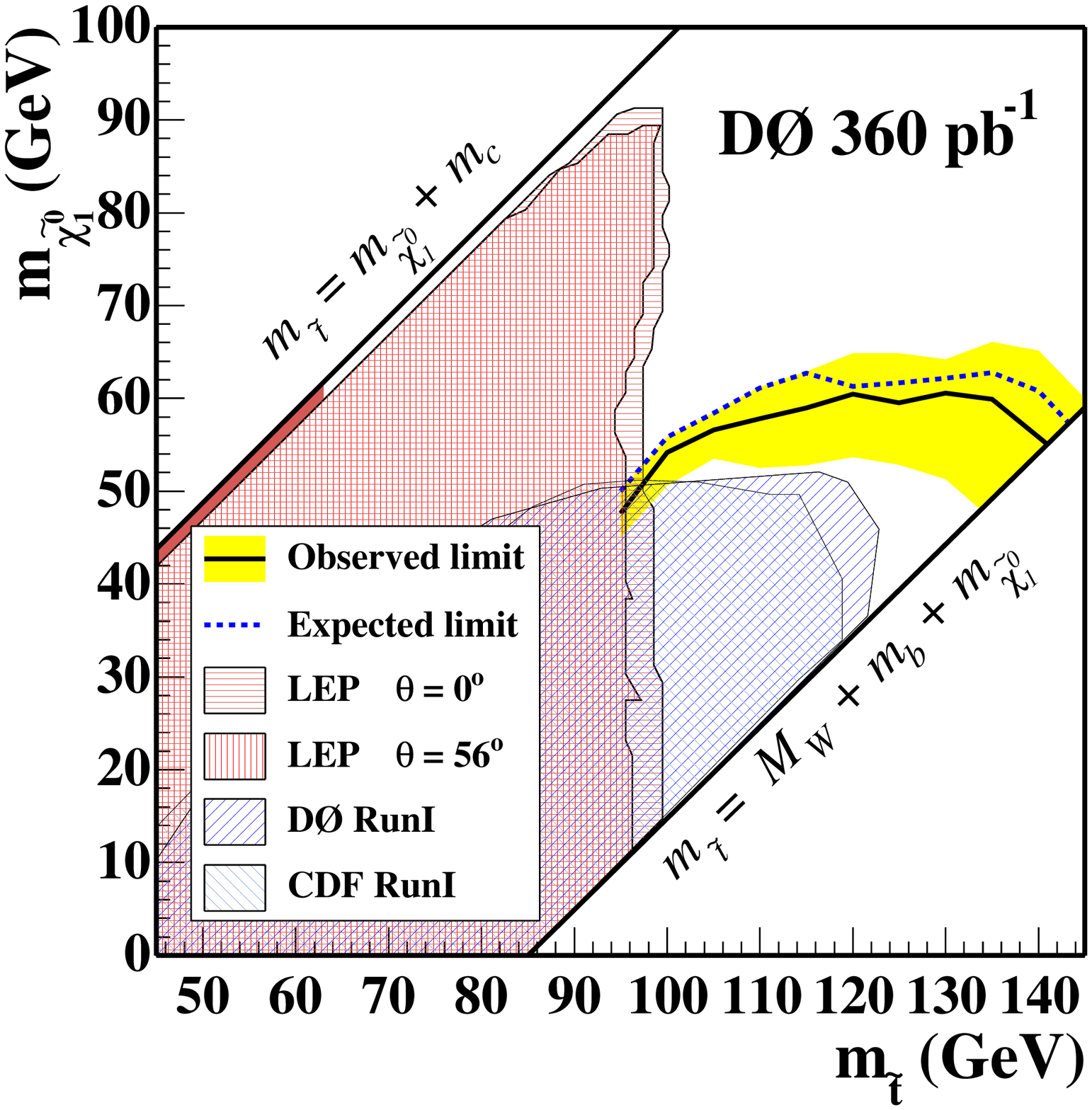}
\caption{\label{exclu}
Domain in the (\mst,\mchi) plane excluded at the 95\% C.L. by the present 
search (region below the solid curve), under the assumption that the stop 
decays exclusively into 
$c\chiz$ and for the nominal production cross section. The expected 
exclusion contour is shown as a dashed curve. The effect of increasing or
decreasing the production cross section by its uncertainty due to the PDF 
and $\mu_{rf}$ choices is indicated for the observed exclusion contour by
the shaded band. Results from previous searches for stop pair production 
in the $\st\to c\chiz$ decay channel are also indicated\,\cite{LEP,CDF,DRunI}.
The dark shaded band at small $\mst-\mchi$ is excluded by Ref.\,\cite{Aleph}.
The LEP results are shown for two values of $\theta$, the mixing angle in the 
stop sector.}
\end{figure}

As can be seen in Table\,\ref{resoptim},
no significant excess of data was observed in any of the optimized selections.
Signal production cross section upper limits were therefore derived with the 
above systematic uncertainties taken into account. Examples are given in 
Table\,\ref{sigeff}, together with the corresponding theoretical cross 
sections. To determine an exclusion domain in the (\mst,\mchi) plane, the
following procedure was used. 
For a given \mst\, the signal confidence level $\mbox{\sl CL}_s$ was computed as a 
function of \mchi\ in the modified frequentist approach\,\cite{CLs}, and the 
95\% C.L. lower limit on \mchi\ was determined as the \chiz\ mass for which 
$\mbox{\sl CL}_s = 5\%$. In this procedure, the theoretical NLO cross sections
predicted by {\sc prospino} with the CTEQ6.1M PDFs were used. The nominal cross
section was obtained for $\mu_{rf}=\mst$. Theoretical uncertainties on the 
stop pair production cross section arise from the choices of PDFs and of
renormalization and factorization scale. The variations observed 
with the CTEQ6.1M error PDF set, as well as the 
changes induced when $\mu_{rf}$ is modified by a factor of two up or down, 
result in a typically $\pm 20$\% change in the theoretical cross section when
combined in quadrature.
The exclusion contour in the (\mst,\mchi) plane thus obtained is shown as a 
solid curve in
Fig.\,\ref{exclu} for the nominal production cross section.
The corresponding expected exclusion contour is shown as a dashed 
curve. The effect of the PDF and scale uncertainties on the observed
exclusion contour is shown as a shaded band.

This analysis, performed under the assumption that the 
stop decays exclusively into a charm quark and the lightest neutralino, 
extends the stop and $\chiz$ mass domain excluded by previous 
experiments\,\cite{LEP,CDF,DRunI}. 
For the nominal stop pair production cross section, the largest 
stop mass excluded is 141\,GeV, obtained for $\mchi= \mst-m_b-m_W = 55$\,GeV. 
Taking into account the theoretical uncertainty on the production cross 
section, the largest stop mass limit is 134\,GeV, obtained for 
$\mchi = 48$\,GeV. 

%
We thank the staffs at Fermilab and collaborating institutions, 
and acknowledge support from the 
DOE and NSF (USA);
CEA and CNRS/IN2P3 (France);
FASI, Rosatom and RFBR (Russia);
CAPES, CNPq, FAPERJ, FAPESP and FUNDUNESP (Brazil);
DAE and DST (India);
Colciencias (Colombia);
CONACyT (Mexico);
KRF and KOSEF (Korea);
CONICET and UBACyT (Argentina);
FOM (The Netherlands);
PPARC (United Kingdom);
MSMT (Czech Republic);
CRC Program, CFI, NSERC and WestGrid Project (Canada);
BMBF and DFG (Germany);
SFI (Ireland);
The Swedish Research Council (Sweden);
Research Corporation;
Alexander von Humboldt Foundation;
and the Marie Curie Program.
%

\end{document}